\documentclass[preprint,12pt]{elsarticle}

\usepackage{amssymb}
\usepackage{amsthm}
\usepackage{amsmath}

\journal{Physics Letters A}

\begin{document}

\begin{frontmatter}



\title{Variational model for one-dimensional quantum magnets}


\author[A1,A2,A3]{Yu. B. Kudasov}
\ead{yu{\_}kudasov@yahoo.com}

\author[A1,A2]{R. V. Kozabaranov}

\address[A1]{Sarov Physics and Technology Institute, NRNU "MEPhI", 6 Dukhov str.,Sarov, 607186, Russia}

\address[A2]{National Research Nuclear University "MEPhI", 31 Kashirskoe ave., Moscow, 115409, Russia}

\address[A3]{Russian Federal Nuclear Center - VNIIEF, 37 Mira ave., Sarov, 607188, Russia}

\begin{abstract}
A new variational technique for investigation of the ground state and correlation functions in 1D quantum 
magnets is proposed. A spin Hamiltonian is reduced to a fermionic representation by the Jordan-Wigner 
transformation. The ground state is described by a new non-local trial wave function, and the total energy is 
calculated in an analytic form as a function of two variational parameters. This approach is demonstrated with 
an example of the XXZ-chain of spin-1/2 under a staggered magnetic field. Generalizations and applications of 
the variational technique for low-dimensional magnetic systems are discussed.  

\end{abstract}


\end{frontmatter}


\section{Introduction}
\label{Intro}

One-dimensional magnetic systems, both a simple chain and complex ones like decorated chains,
zig-zag and ladder structures, are drawn a considerable attention of theoreticians and experimentalists 
\cite{mattis,takahashi,nunner}.
It is related to recent progress in the synthesis of one-dimensional molecular magnets \cite{bogani} and 
quasi-one-dimensional magnetic structures in crystalline substances \cite{vasilev}.  

A Heisenberg chain of spin-1/2 is one of the most fundamental and thoroughly investigated models of magnetism
\cite{mattis}. Nevertheless, a few exotic phases were recently revealed: the ground state with E8 symmetry 
under a transverse magnetic field in CoNb$_2$O$_6$ \cite{coldea}, Bose glass in (Yb$_{1-x}$Lu$_x$)$_4$As$_3$ 
\cite{kamieniarz},
and etc. 

Analytic solutions for the Heisenberg antiferromagnetic (AFM) chain with longitudinal magnetic field
are well-known, namely the ground state energy \cite{mattis,takahashi,griffiths} and excitation spectrum  
\cite{cloizeaux,faddeev} which is gapless at magnetic fields below the critical value \cite{johnson}.
At the same time, a spin gap is observed in various one- and quasi-one-dimensional magnets \cite{vasilev}. In 
some cases the gap stems from a staggered magnetic field appeared due to the Dzyaloshinskii-Moriya interaction 
\cite{oshikawa} or an effect of the transverse magnetic field on an anisotropic zig-zag chain \cite{bogani}. 

There are no analytic solutions for the Heisenberg chain under the staggered magnetic field. An asymptotic 
solution for the
isotropic chain in the limit of weak staggered field ($h_{st}\rightarrow 0$) was obtained by transformation to 
the sine-Gordon model. It is valid within a very narrow region
in the vicinity of $h_{st}=0$ \cite{affleck}. The finite-temperature density-matrix
renormalization group (DMRG) theory allowed resolving the problem at wider range of finite staggered field 
\cite{shibata}. 
Recently the XXZ-chain with staggered magnetic field was thoroughly investigated using the
mean-field approach with fluctuation corrections up to the second order and the exact diagonalization 
on finite clusters \cite{paul}. The Heisenberg Hamiltonian was preliminarily mapped onto
a fermionic representation by the Jordan-Wigner transformation \cite{jordan}. 
It was shown that the mean-field approximation
with the corrections in a number of cases gives unsatisfactory results. In particular, in the limit 
$h_{st}\rightarrow 0$
the ground state energy of the XY-chain diverges and the spin gap does for the isotropic Heisenberg
chain \cite{paul}.

On the other hand, the mapping on the fermionic representation by means of the Jordan-Wigner
transformation makes possible applying well-developed techniques of strongly correlated
Fermi systems theory. In particular, a variational Gutzwiller approach \cite{gutzwiller} 
has allowed calculating
the ground state energy of the Hubbard model for the infinite-dimensional lattice. It was
also successively applied to low dimensional lattices up to one-dimensional chain \cite{gebhard}. 
The Gutzwiller trial wave function had been intended for control of intrasite correlations, its
generalization enabled to include non-local correlations between the nearest neighbors \cite{kudasov}.
It was shown that this trial wave function produces a good approximation of the ground state for the
Hubbard model even in the one-dimensional case. Since the fermionic representation for Heisenberg
chain contains interactions between the nearest-neighboring sites, the generalized non-local trial wave
function seems to be a promising candidate for its ground state description. 

In the present Letter, we propose a new variational approach to one-dimensional quantum magnets
and illustrate it by example of the Heisenberg XXZ-chain with the staggered magnetic field. The procedure 
includes the follow steps: (i) the transition to the fermionic representation by means of the
Jordan-Wigner transformation, (ii) development of the trial wave function for spinless fermions,
(iii) calculation of the ground state energy, correlation functions, and other characteristics 
with the trial wave function.

\section{Jordan-Wigner transformation}
\label{JW}

The Hamiltonian of spin-1/2 Heisenberg XXZ chain under the staggered magnetic field has the following
form \cite{paul}
\begin{eqnarray}
\hat{H}=\hat{H}_{xy}+\hat{H}_{zz}+\hat{H}_{st}
\label{H}
\end{eqnarray}
where $\hat{H}_{xy}=\frac{J}{2}\sum_{i}^{N}\left( 
\hat{S}_{i}^{+}\hat{S}_{i+1}^{-}+\hat{S}_{i}^{-}\hat{S}_{i+1}^{+}\right) $ and
$\hat{H}_{zz}=J\Delta\sum_{i}^{N}\hat{S}_{i}^{z}\hat{S}_{i+1}^{z}$ are the $xy$- and $zz$-terms of
the Hamiltonian,
$\hat{H}_{st}=h_{st}\sum_{i}^{N}\left( -1\right) ^{i}\hat{S}_{i}^{z}$ is the contribution of
the staggered magnetic field 
$h_{st}$, $\hat{S}_{i}^{+}\left(\hat{S}_{i}^{-} \right)$ and $\hat{S}_{i}^{z}$ are the operators of
spin raising (lowering) and its component along the $z$-axis. The constant $J$ is assumed to be positive.
Below we discuss mainly behavior of the anisotropic AFM chain ($\Delta \geq 0$), however the results remain 
valid for the ferromagnetic (FM) chain also ($\Delta < 0$).

The Jordan-Wigner transformation allows representing the spin operators through creation (annihilation)
operators for spinless fermions at the $i$-th chain site $\hat{c}_{i}^{\dagger}$~($\hat{c}_{i}$) 
\cite{jordan}:   
\begin{align}
\hat{S}_{i}^{+}&=\hat{c}_{i}^{\dagger}\exp\left( i\pi\sum_{j=1}^{i-1}\hat{n}_{j} \right), \nonumber \\
\hat{S}_{i}^{-}&=\exp\left(-i\pi\sum_{j=1}^{i-1}\hat{n}_{j}\right)\hat{c}_{i}, \label{JW} \\
\hat{S}_{i}^{z}&=\hat{n}_{i}-1/2.  \nonumber
\end{align}
This reduces the Hamiltonian (\ref{H}) to a model of fermionic chain \cite{paul}:
\begin{align}
\hat{H} &= \hat{H}_{0}+\hat{H}_{1},\label{fc}\\
\hat{H}_{0}&=\sum_{j}^{N/2}\left[ \frac{J}{2}\left( 
\hat{a}_{j}^{\dagger}\hat{b}_{j}+\hat{a}_{j+1}^{\dagger}\hat{b}_{j}+h.c.\right) \right. \nonumber \\
& \left. + h_{st}\left( \hat{a}_{j}^{\dagger}\hat{a}_{j}-\hat{b}_{j}^{\dagger}\hat{b}_{j} \right) \right], 
\nonumber \\
\hat{H}_{1}&= J\Delta\sum_{j}^{N/2}\left[ \left( \hat{a}_{j}^{\dagger}\hat{a}_{j}-\frac{1}{2}\right) \left( 
\hat{b}_{j}^{\dagger}\hat{b}_{j}-\frac{1}{2} \right) \right. \nonumber \\
&\left. + \left( \hat{b}_{j}^{\dagger}\hat{b}_{j}-\frac{1}{2}\right). \left( 
\hat{a}_{j+1}^{\dagger}\hat{a}_{j+1}-\frac{1}{2}\right) \right], \nonumber 
\end{align}
where $\hat{a}_{i}^{\dagger}$ and $\hat{b}_{i}^{\dagger}$ are the creation operators for spinless fermions at 
the
$A$ and $B$ sublattices correspondingly, that is, $\hat{a}_{i}^{\dagger} \equiv \hat{c}_{i}^{\dagger}$ ($i \in 
A$) and $\hat{b}_{j}^{\dagger} \equiv \hat{c}_{j}^{\dagger}$ ($j \in B$). 
The Hamiltonian (\ref{fc}) contains quadratic ($\hat{H}_{0}$) and biquadratic ($\hat{H}_{1}$) parts.
The first one corresponds to the kinetic energy of a tight-binding model, and the second represents
an interaction between the fermions at the nearest-neighboring chain sites.

The quadratic part of the Hamiltonian $\hat{H}_{0}$ is diagonalized by a
unitary transformation \cite{paul}
\begin{eqnarray}
\hat{H}_{0d}=U\hat{H}_{0}U^{-1}=\sum_{k}\varepsilon_{k}\left( 
\hat{\alpha}_{k}^{\dagger}\hat{\alpha}_{k}-\hat{\beta}_{k}^{\dagger}\hat{\beta}_{k}\right)
\label{U}
\end{eqnarray}
where
$\varepsilon_{k}=\sqrt{J^{2}\cos^{2}({k/2})+h_{st}^{2}}$. Hereinafter we use a reduced Brillouin zone
corresponding to the doubled chain period, that is, $k/2 \rightarrow k$. It should be mentioned that
$\hat{H}_{0d}$ corresponds to the XY-model with the staggered magnetic field. In the ground state,
the branch with the negative eigenvalues is fully filled up ($n_{\beta k} = 
\hat{\beta}_{k}^{\dagger}\hat{\beta}_{k} =1$), and that with the positive ones is empty
($n_{\alpha k} = \hat{\alpha}_{k}^{\dagger}\hat{\alpha}_{k} =0$). Thus the ground state of the
XY-chain in the staggered magnetic field is determined exactly:  
\begin{eqnarray}
|{\tilde{\varphi}}\rangle=\prod_{k}\beta_{k}^{\dagger}|0\rangle.
\label{gs}
\end{eqnarray}

For the sake of convenience, below we apply a representation of the operators  $\hat{a}_{i}^{\dagger}$
and $\hat{b}_{i}^{\dagger}$ expressed in terms of the diagonal operators $\hat{\alpha}_{k}^{\dagger}$ and 
$\hat{\beta}_{k}^{\dagger}$ by means of the inverse transformation 
$|\varphi\rangle=U^{-1}|\tilde{\varphi}\rangle$.

\section{Trail wave function}
\label{wf}

To generate a non-local trial wave function one should define projection operators on all possible
configurations of the nearest neighboring pairs of sites in the chain \cite{kudasov}. There are four
such configurations for spinless fermions 
\begin{align}
\hat{Y_{1}}&=\sum_{<i,j>}\left( 1-\hat{n}_{i}^{A}\right) \left( 1-\hat{n}_{j}^{B}\right), \nonumber \\
\hat{Y_{2}}&=\sum_{<i,j>}\hat{n}_{i}^{A}\left( 1-\hat{n}_{j}^{B}\right),  \nonumber \\
\hat{Y_{3}}&=\sum_{<i,j>}\left( 1-\hat{n}_{i}^{A}\right) \hat{n}_{j}^{B},  \nonumber \\
\hat{Y_{4}}&=\sum_{<i,j>}\hat{n}_{i}^{A}\hat{n}_{j}^{B} 
\label{y}
\end{align}
where $<...>$ denotes a sum over all the pairs of the nearest neighbors. The sites in the pairs
belong to different sublattices:
($i \in A$) and ($j \in B$). It is worth noticing that the operators $\hat{Y_{k}}$ are not completely
independent \cite{kudasov}. If we consider average values normalized to a single chain site 
$y_k=L^{-1}<\hat{Y_{k}}>$, which can be interpreted as probabilities of the corresponding configurations, they 
turn out to be related one another
by conditions of normalization ($\sum_k {y_{k}}=1$) and half-band filling ($y_{2}+y_{3}+2 y_{4}=1$).
Thus it is convenient to introduce a pair of independent symmetrized operators 
$\hat{M}=\hat{Y_{3}}-\hat{Y_{2}}$ and $\hat{P}=\hat{Y_{3}}+\hat{Y_{2}}$. Their physical meaning
can be clarified by the averages $m=L^{-1}<\hat{M}>$ and $p=L^{-1}<\hat{P}>$: the limiting value
$m=1$ corresponds to the chain state when all the site of the B sublattice are filled up ($n_B=1$) and
all the sites of the A sublattice are empty ($n_A=0$), and in the opposite limit ($m=-1$) vice versa
($n_A=1$ and $n_B=0$). That is why, $m$ denotes the AFM magnetization. The other average $p$ defines
a nearest-neighbor spin-spin correlation function
\begin{equation}
\langle \hat{S}_{i}^{z} \hat{S}_{i+1}^{z} \rangle = (1-2p)/4.
\label{sc}
\end{equation}

The trial wave function of Gutzwiller's type with variational parameters corresponding
to the configurations of pairs of the nearest neighboring sites takes the form
\begin{eqnarray} 
|\psi\rangle=g_{p}^{\hat{P}} g_{m}^{\hat{M}} |\psi_{0}\rangle
\label{tf}
\end{eqnarray}
where $g_{p}$ and $g_{m}$ are the nonnegative variational parameters and
$|\psi_{0}\rangle$ is the initial many-body wave function which can be
represented by the exact wave function for noninteracting fermions 
(\ref{gs}). It can be shown that the transformation (\ref{tf}) 
retains the permutation antisymmetry of the initial wave function as well
its point and translational symmetries \cite{kudasov}. 
On the other hand, one can expand the initial wave function as a series in configurations 
$|\psi_{0}\rangle=\sum_{\Gamma}A_{\Gamma}|\Gamma\rangle$ where $A_{\Gamma}$ is the
complex amplitude of the configuration $|\Gamma\rangle$. Then it becomes clear that weights of 
the configuration  amplitudes in the trial wave function are modified depending on
the number of particular arrangements of nearest neighboring pairs, that is, 
$|\psi_{0}\rangle=\sum_{\Gamma}A_{\Gamma}g_{p}^{P_\Gamma} g_{m}^{M_\Gamma} |\Gamma\rangle$ where
$P_{\Gamma}=\langle\Gamma|\hat{P}|\Gamma\rangle$, $M_{\Gamma}=\langle\Gamma|\hat{M}|\Gamma\rangle$. 
For example, if $g_p >1$ the weight of configurations rises with increase of the number of
nearest-neighbor pairs with a single fermion. Thus, one can control nonlocal correlations
(intensify or supress) in the trial wave function. It should be mentioned that the trial wave function in the 
form (\ref{tf}) is nonnormalized.

It is convenient to accept the exact solution for noninteracting fermions at the zero magnetic field
$|\psi_{0}(0)\rangle$ ($\Delta=0$, $h_{st}=0$) as the initial wave function. To make the procedure
more flexible one can use a set of initial wave functions for noninteracting fermions
$|\psi_{0}(h_e)\rangle$ under an effective field $h_e \rightarrow h_{st}$ which does not 
necessarily coincide with the staggered magnetic field.

In the limit of large number of particles $N$ a distribution of configurations number on
$M_\Gamma$ and $P_\Gamma$ has a sharp maximum. The maximum width is of the order of $N^{-1/2}$ with an 
exponential decay while going away from the maximum. That is why, one can limit oneself to the configurations 
with the weight $R(m,p)=W(m,p) g_{m}^{2Lm}  g_{p}^{2Lp}$ close to the maximal one
where $W(m,p)$ is the number of configurations with $M_\Gamma=mL$ and $P_\Gamma=pL$. This function is 
evaluated by pseudo-ensemble technique  \cite{kikuchi,ziman,kudasov}.
A necessary condition of the maximum is determined by the following equations \cite{kudasov} 
\begin{eqnarray} 
\frac{\partial \ln [R(m,p)]}{\partial m}=0,\, \frac{\partial \ln [R(m,p)]}{\partial p}=0.
\label{max}
\end{eqnarray}
The equations (\ref{max}) lead to expressions for $g_p$ and $g_m$:
\begin{eqnarray} 
g_p=\frac{(p^2-m^2)^\frac{1}{4}}{(1-p)^\frac{1}{2}}, 
g_m=\left[\frac{(1-m)(p+m)}{(1+m)(p-m)}\right]^{\frac{1}{4}}.
\label{gg}
\end{eqnarray}

\section{The ground state energy}
\label{ge}

The total ground state energy of the system per lattice site 
as a function of the variational parameters $m$ and $p$ may be written as \cite{gutzwiller,kudasov}
\begin{eqnarray}
E=q'\langle\epsilon_{k}\rangle_{0}-\frac{m h_{st}}{2}+\frac{J\Delta}{4}\left( 1-2p\right), 
\label{E}
\end{eqnarray}
where $q'=\frac{q}{q_{0}}$ and $q=L^{-1} \sum_{<ij>}(\hat{a}_{i}^{\dagger}\hat{b}_{j}+H.c.)$ is
the first-order density function, and the normalizing factor $q_{0}$ is its value for noninteracting
fermions, i.e.  $\Delta=0$. The kinetic energy of noninteracting fermions $\langle\epsilon_{k}\rangle_{0}$ is 
calculated from
the dispersion relation $\varepsilon_{k}$:
\begin{eqnarray}
\langle\epsilon_{k}\rangle_{0}=-\frac{1}{\pi}\int_{0}^{\pi/2}\frac{J^{2}\cos^{2}k}{\sqrt{J^{2}\cos^{2}k
+h_{e}^{2}}}dk,
\label{e0}
\end{eqnarray}
from which we obtain
\begin{eqnarray}
\langle\epsilon_{k}\rangle_{0}=-\frac{1}{\pi} \left[\sqrt{h_e^2+J^2} 
\mathsf{E}\left(\frac{J}{\sqrt{h_e^2+J^2}} \right) \right.
\nonumber \\
- \left. \frac{h_e^2 \mathsf{K} \left( \frac{J}{\sqrt{h_e^2+J^2}} \right)}{\sqrt{h_e^2+J^2}} \right],
\label{e0}
\end{eqnarray}
where $\mathsf{K}$ and $\mathsf{E}$ are the complete elliptic integral of the first and second order
correspondingly. It also should be mentioned that $h_e$ and $m$ are unambiguously related to one
another as follows
\begin{eqnarray}
m=\frac{2h_e}{\pi \sqrt{h_e^2+J^2}} \mathsf{K} \left[ \frac{J}{\sqrt{h_e^2+J^2}} \right].  
\label{m}
\end{eqnarray}

The function $q$ expresses the change of the density function with variational parameters. It can
be calculated by means of technique developed earlier \cite{gutzwiller,kudasov}.
For instance, a fraction of configurations, in which the site $i$ is filled up and $j$ is empty, is $y_2$.
After a transition of the fermion from the site $i$ to $j$ the average value $<\hat{M}>$
increases by 4 that leads to a multiplier $g_{m}^4$. In addition, one should take into account that
configurations of the pairs adjacent to $ij$ also alters with the transition. A detailed discussion
of the computation technique will be presented elsewhere. Collecting all the terms together
we obtain 
\begin{eqnarray}
q=y_{2}g_{m}^4\left(\frac{y_{2}g_{p}^{-1}+y_{4}g_{p}}{y_{2}+y_{4}} \right)\left(\frac{y_{1}g_{p}+y_{2} 
g_{p}^{-1}}{y_{1}+y_{2}} \right)+\nonumber \\
+\frac{y_{3}}{g_{m}^{4}}\left( 
\frac{y_{1}g_{p}+y_{3}g_{p}^{-1}}{y_{1}+y_{3}}\right)\left(\frac{y_{3}g_{p}^{-1}+y_{4}g_{p}}{y_{3}+y_{4}}. 
\right)
\label{q}
\end{eqnarray}
Here we express $g_m$ and $g_p$ through $m$ and $p$ using formulae (\ref{gg}) and replace $y_k$ by
$y_1=y_4=(1 - p)/2$, $y_2=(p-m)/2$ and $y_3=(p+m)/2$. Then $q$ becomes a function of $m$ and $p$.
A substitution of (\ref{q}), (\ref{m}) and (\ref{e0}) into (\ref{E}) gives the total energy as a function of 
the variational parameters  $m$ and $p$ in an analytic form. The ground state energy
is determined by its numerical minimization. 

\begin{figure}
	\includegraphics[width=0.95\textwidth]{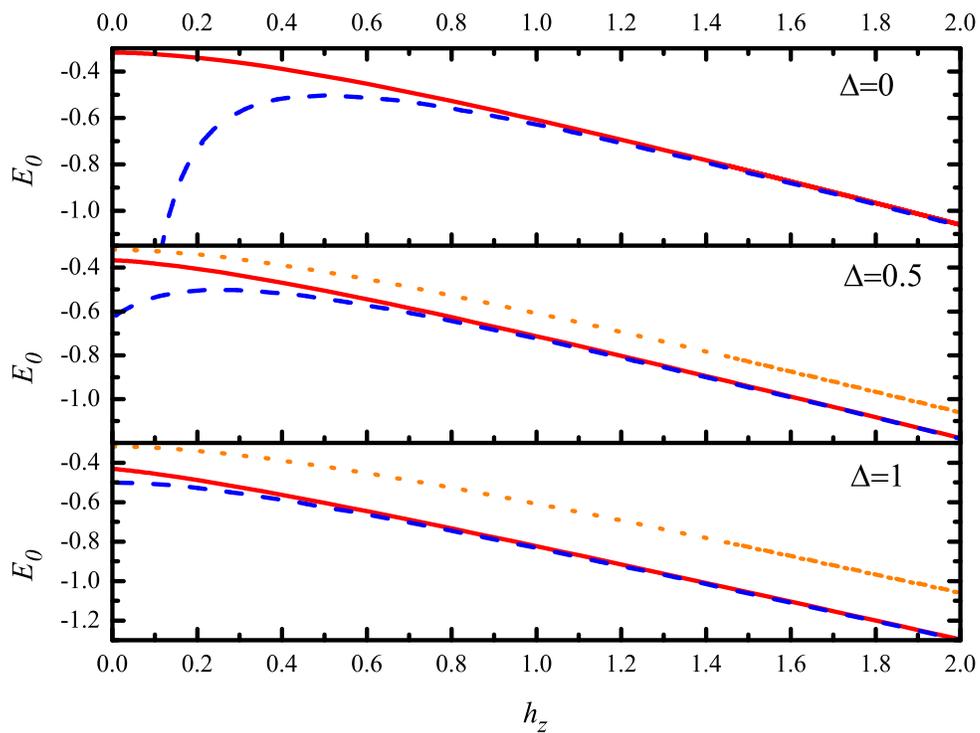}
	\caption{\label{f1} The ground state energy: results of the present work (the solid red line),  the 
mean-field theory with the second-order fluctuation corrections \cite{paul} (the blue dash line), the exact 
solution of the XY-model (the orange dotted line). }
\end{figure}

Results of the ground state energy calculation for the Heisenberg XXZ-chain under the staggered magnetic field 
are
shown in Fig.~\ref{f1} for different values of the anisotropy parameter $\Delta$.
They are compared there with the exact solution for the XY-chain and results of the mean-field
theory with the fluctuation corrections \cite{paul}. Our variational solution coincides with the exact
one at $\Delta=0$ whereas the corrected mean-field solution diverges at $h_{st}=0$ \cite{paul}. In the limit 
of large values of $h_{st}$ and $\Delta$ the system tends to the N\'{e}el state and both the solutions 
coincide. In case of the isotropic AFM chain ($\Delta=1$) the ground state energy was estimated as
$E_0=-0.4311 J$ at $h_{st}=0$. This value differs from the exact one obtained by the Bethe ansatz 
($(1/4-\ln 2)J\approx-0.4431 J$) by 2.7~\%. 

While considering the FM exchange ($\Delta < 0$) at $h_{st}=0$ an additional minimum of the total
energy at $p \rightarrow 0$, $m \rightarrow 0$ corresponding to the FM state is revealed. 
It becomes the global one below $\Delta=-1.075$. This is close to the exact value $\Delta=-1$.
It should be pointed out that for a rigorous description of the FM state it is necessary to go beyond the 
half-band filling assumption because
the saturated FM state correspond to totally filled up or empty band in the fermion
representation. An additional variational parameter appears in this case.

\begin{figure}
	\includegraphics[width=0.95\textwidth]{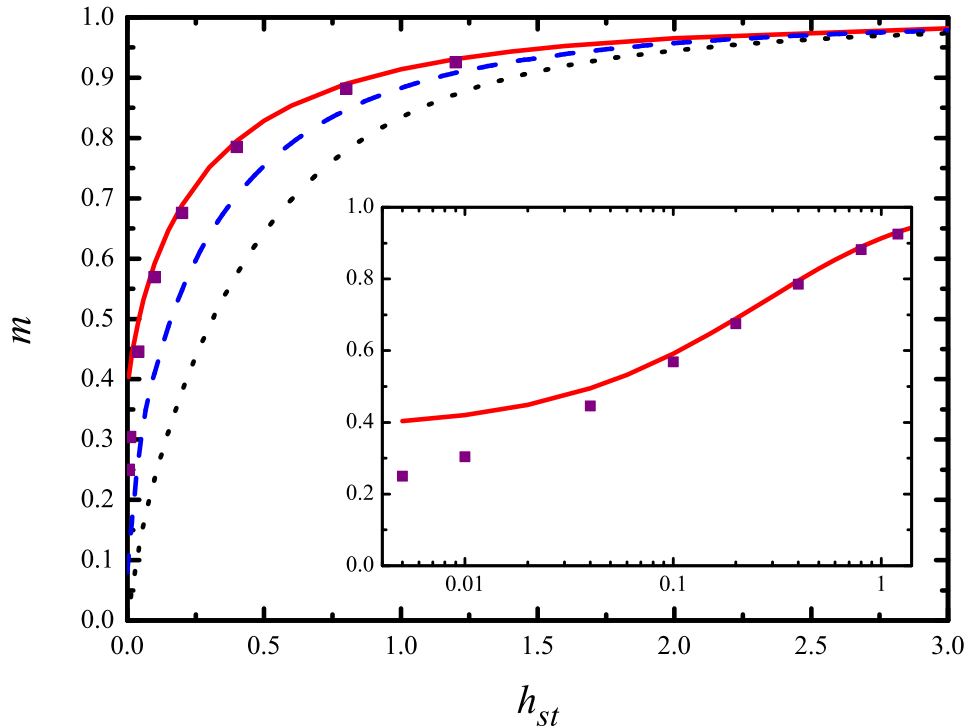}
	\caption{\label{f2} The AFM magnetization $m$ as a function of the staggered magnetic field $h_{st}$. The 
dotted line is the solution of the XY-model ($\Delta=0$), the dash line corresponds to
	 $\Delta=0.5$, and the solid line does to the isotropic Heisenberg model
		($\Delta=1$). The solid squares denotes the solution by the DMRG technique at $\Delta=1$ 
\cite{shibata}. The enlarged fragment of the function at weak magnetic fields is shown in the insert.}
\end{figure}

The AFM magnetization at the ground state 
as a function of the staggered magnetic field is shown in Fig.~\ref{f2}.
Comparison of the obtained solution with the DMRG theory for the isotropic chain demonstrates a good
agreement everywhere except a narrow region in the vicinity of $h_{st}=0$ as one can see in the insert to
Fig.~\ref{f2}. For $\Delta=0$ the AFM magnetization coincides with the exact solution for the XY model. 

\section{Discussion and conclusion}
\label{dc}

The proposed approach may be applied to quantum one-dimensional magnets including systems with a complex
structure. A necessary condition for its implementation is a reduction of an initial model to a fermionic 
representation by the Jordan-Wigner transformation. In particular, a two-leg Heisenberg ladder
is reduced to a fermionic chain partitioned between two sublattices \cite{nunner} that is very similar to the
present approach.

One may consider the procedure proposed above as a two-component mean-field approach where the first
component determines the AFM magnetization ($m$) and the second one controls the spin-spin
correlation function (\ref{sc}).

The ground state obtained by the variational approach is exact by definition at $\Delta=0$ and
goes asymptotically to the exact solutions in the FM and AFM Ising limits. The most sizable divergence
with the exact solution appears for the isotropic chain in the vicinity of zero staggered magnetic field
($\Delta=1$, $h_{st}\rightarrow0$). The total energy $E(m,p)$ in this case becomes a flat function close to
the global minimum. That is why, small variations in the energy correspond to large shifts of the minimum.
From the physical point of view, one can see that the correlation length increases approaching to $\Delta=1$
in the XY region and the susceptibility to the staggered field unrestrictedly grows.

In the present work, a trial wave function with nonlocal projection operators restricted to nearest-neighbor 
pairs was used. In the framework of proposed approach it is possible to extend the correlations in the trail 
wave function up to 3 or 4 adjacent chain sites. The number of independent variational parameters has to be 
increased
up to 5 or 7. It was shown in Ref.~\cite{kudasov} by example of the Hubbard model that the technique
remains efficient in this case. The extended trial wave function should improve the description of
the ground state in the vicinity of $\Delta=1$, $h_{st}=0$.   




	
	

\end{document}